\documentclass{article}

\usepackage[english]{babel}
\usepackage[a4paper, margin=2.54cm]{geometry}
\usepackage{authblk}
\usepackage{amsmath}
\usepackage{amssymb}
\usepackage{graphicx}
\usepackage{float}
\usepackage{xurl}
\usepackage[colorlinks=true, allcolors=blue]{hyperref}
\usepackage{caption}
\usepackage{titling}
\AtBeginDocument{
    
    \captionsetup{labelsep=space, labelfont=bf}
}

\title{Building atomistic models of heterointerfaces with optimal transport}

\author{Yuxuan Tang}
\author{Keith T. Butler\thanks{Corresponding author: \href{mailto:k.t.butler@ucl.ac.uk}{k.t.butler@ucl.ac.uk}}}
\affil{Department of Chemistry, University College London, London, WC1E 6BT, United Kingdom}

\date{}

\begin{document}
\maketitle

\begin{abstract}

Heterogeneous interfaces underpin technologies from microelectronics to energy conversion and storage, but their configurational complexity precludes exhaustive first-principles screening of interface registries. Although data-driven approaches can alleviate this burden, they remain limited by sparse interface datasets. Here, we introduce an energy-independent workflow that represents coherent interfaces as attributed graphs, quantifies their similarity to parent bulk environments using the fused Gromov-Wasserstein (FGW) distance, and couples this metric with Bayesian optimization over the in-plane registry space. We assess the approach for KI/NaCl, GaP/GaAs and GaN/\(\mathrm{Al_{2}O_{3}}\) interfaces spanning ionic, covalent and mixed-bonding regimes, using hierarchical validation with MACE and density functional theory (DFT). Comparison with single-point energy landscapes shows that the FGW distance captures registry-dependent periodicity, while interfaces exhibit deviations between structural and energetic extrema, reflecting additional chemistry-specific contributions. Furthermore, FGW distances show an overall association with relaxed energies. Under limited screening budgets, FGW-guided registry selection consistently outperforms random search and is more robust across interface systems than selection guided by pretrained MACE energies. The workflow converts the qualitative notion of bulk-like continuity into a quantitative prescreening criterion, enabling efficient registry exploration and providing physically informed candidate structures for materials discovery workflows.

\end{abstract}

\section{Introduction}

Heterogeneous interfaces are ubiquitous in modern materials systems, underpinning technologies ranging from microelectronics to energy conversion and storage\cite{XU2024, Zhang2022, Fechete2012, Gao2017, Luo2021, Shao2022, Hui2020, An2025}. By coupling distinct materials, heterointerfaces give rise to emergent properties that are absent in the corresponding bulk phases, thereby governing device performance and stability\cite{Zhong2017, Majchrzak2025}. This pronounced interfacial sensitivity has driven extensive research efforts to elucidate, control and rationally design heterointerfaces.

High-performance heterogeneous interfaces have been synthesized and tuned through advanced thin-film growth and interface engineering\cite{Wang2024, Bansal2011, Thapa2021}. However, the structural disorder, chemical inhomogeneity and dynamic reconstruction that may arise at interfaces, together with their buried nature make exhaustive experimental characterization prohibitively difficult. Atomistic simulations therefore provide an indispensable route to systematically probe interfacial structures and structure-property relationships. Among these methods, periodic density functional theory (DFT) remains a well-established approach for evaluating interfacial structures, but its high computational cost limits the number of candidate interfacial configurations that can be assessed\cite{Zhuo2018, Butler2019}.

Recent efforts have increasingly turned to automated and machine-learning-assisted workflows. Surrogate models trained on first-principles datasets can rapidly predict principal interfacial properties, such as ionization potentials\cite{Choudhary2024} and adsorption energies\cite{Ulissi2017}, enabling high-throughput screening across a large materials space. Beyond surrogate-based screening, machine-learned interatomic potentials (MLIPs) have been widely adopted to expedite atomistic simulations by approximating the underlying potential-energy surface from DFT data. While significantly extending accessible system sizes and time scales, MLIPs enable structural relaxation and molecular dynamics at near DFT accuracy\cite{Mouvet2022, Lin2024}. More recently, rather than explicitly traversing the potential-energy landscape, generative models provide a complementary, data-driven route that directly proposes candidate atomic arrangements from learned distributions\cite{Kim2020, Zeni2025, Gao2025}. Collectively, these machine learning (ML) approaches have substantially expanded the accessible scope of interfacial configuration space.

Despite these advantages, the effectiveness of current ML strategies depends strongly on the availability of informative prior data, suitable structural representations, and tractable candidate-generation schemes\cite{walker2026carbon}. For bulk materials, the reference data are increasingly accessible owing to the availability of large, standardized materials databases, such as the Materials Project (MP)\cite{Jain2013} and the Open Quantum Materials Database (OQMD)\cite{Saal2013}. For heterogeneous interfaces, in contrast, high-quality reference data remains scarce, and most existing studies adopt simulation-driven, system-specific workflows with on-the-fly data generation\cite{Hrmann2025}. In parallel, multiple surface orientations, terminations and interfacial registries, as well as possible intermixing, defects and reconstruction, make the configurational space of interfaces far more complicated than that of bulk materials. A variety of geometric matching schemes and descriptor-based approaches have been developed to navigate this configurational complexity \cite{Moayedpour2021, Taylor2025}. However, they generally rely on simplified structural representations, empirically chosen parameters, and case-dependent assumptions, limiting their generality and transferability. Related strategies have also used local-environment descriptors and structural similarity to accelerate the screening of complex atomistic configuration spaces\cite{zhang-rematch}. A general physical principle is that interface configurations preserving local structural continuity with the constituent bulk phases tend to be physically reasonable and energetically favourable starting points for interface modelling\cite{Xiong2017}. Nevertheless, this physically intuitive heuristic has not been formulated as a unified and quantitative criterion that can be systematically applied across different interface systems.

Atomic structures are increasingly represented as graphs, in which nodes encode local chemical information and edges encode geometric relations\cite{Xie2018, Reiser2022, Merchant2023}. Within this representation, structural comparison is cast as a graph similarity problem. Unlike other optimal transport methods such as the Wasserstein distance, which operates on feature distributions in a shared space, and the Gromov-Wasserstein distance, which compares relational structures but neglects node attributes\cite{Mmoli2011}, the fused Gromov-Wasserstein (FGW) distance\cite{Vayer2019} provides a principled framework for comparing attributed graphs by jointly accounting for node features and pairwise relational structure\cite{walker2026discovering}. This makes FGW distance well suited to quantifying graph-level similarity between atomistic structures with differences in both geometry and composition.

In this work, we introduce a general FGW-guided workflow that integrates coherent interface construction with Bayesian optimization (BO) over the in-plane registry space. The proposed framework is evaluated on KI/NaCl, GaP/GaAs and GaN/\(\mathrm{Al_{2}O_{3}}\) interfaces, spanning ionic, covalent and mixed-bonding regimes. The interfaces are denoted in film/substrate order and assessed using a hierarchical validation strategy combining large-scale simulations with a pretrained MACE model and high-accuracy DFT calculations. Comparing with MACE single-point energy landscapes, the FGW distance reproduces the registry-dependent periodicity. The structural and energetic extrema are not aligned for the KI/NaCl interface, suggesting additional contributions from long-range electrostatic interactions. MACE and DFT relaxations reveal an overall association between FGW distance and relaxed energy. Under limited budgets, FGW-guided selection identifies low-energy registries more efficiently than random search and shows greater consistency across interface systems than MACE-guided selection. Overall, the FGW metric and workflow establish a well-defined and transferable framework for exploring interfacial configuration space prior to explicit energy evaluation, enabling more systematic and efficient interface modelling.

\section{Results and discussion}

\subsection{FGW-based interface-registry screening}

\begin{figure}[H]
    \centering
    \includegraphics[width=\textwidth, trim=1.5cm 0cm 1.5cm 0cm, clip]{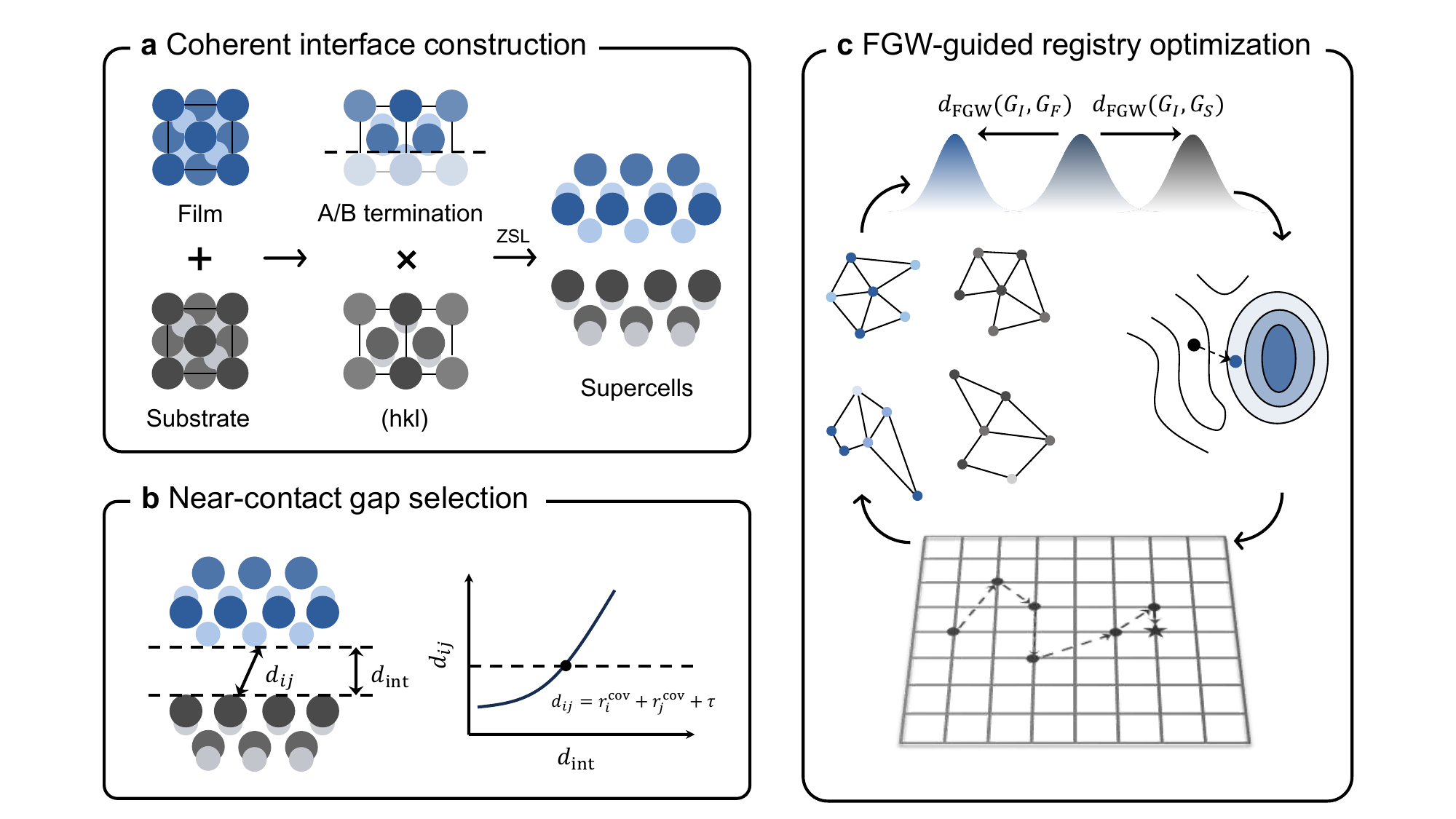}
    \caption{
    Schematic illustration of the FGW-guided workflow for interface modelling. (a) Coherent interface construction via surface enumeration and lattice matching. (b) Near-contact gap determination based on interfacial atomic distances. (c) FGW-guided optimization in the in-plane registry space.
    }
    \label{fig: workflow}
\end{figure}

As illustrated in Fig.~\ref{fig: workflow}, the workflow begins with the conventional unit cells of the film and substrate phases, from which surface orientations and terminations are enumerated and lattice-matched to construct coherent interface supercells. These supercells provide controlled, computationally tractable reference models because the two constituent slabs share a common in-plane lattice periodicity, allowing different registries to be compared at fixed orientation, termination, stoichiometry, imposed strain state and interfacial area. Although semi-coherent and incoherent interfaces can in principle be approximated by large commensurate supercells, these models include additional structural degrees of freedom, such as misfit dislocations, long-range strain fields, and spatially varying local registries, obscuring the isolated effect of in-plane registry and complicating subsequent high-fidelity investigation\cite{Uberuaga2019, Shao2018, Wang2023}.

Before registry screening, a near-contact gap is defined to place the two slabs in a physically meaningful pre-relaxation contact regime. The gap is small enough to generate cross-interface atomic neighbourhoods in the graph representation, and sufficiently large to reduce the influence of  unphysical atomic overlap on the FGW metric. In this way, the computed FGW distance reflects registry-dependent local coordination and geometric continuity rather than artefacts arising from an arbitrary interface gap (full details given in the Methods section). At a fixed near-contact interfacial separation, BO is performed over the two-dimensional registry space using the combined FGW distance as the objective function, defined as the sum of the FGW distances from the candidate interface graph to the film and substrate bulk graphs (see the Methods section for details). This definition is based on the premise that interface-bulk structural similarity provides a reliable proxy for interface compatibility. Structurally compatible registries are therefore identified for each coherent interface supercell.

Fig.~\ref{fig: workflow}a illustrates the construction of interfaces implemented using pymatgen\cite{Ong2013}. Surface orientations and terminations are first enumerated for the film and substrate bulk structures, followed by in-plane lattice matching using the Zur-McGill scheme (ZSL)\cite{Zur1984}. Candidate lattice matches are screened to retain computationally tractable supercells while limiting lattice mismatch and artificial strain.

As shown in Fig.~\ref{fig: workflow}b, the near-contact interfacial separation is determined from cross-interface interatomic distances analysed using pymatgen and the Atomic Simulation Environment (ASE)\cite{HjorthLarsen2017}. Contacts are identified by comparing interatomic distances with the sums of element-specific covalent radii. To avoid dependence on a single arbitrary alignment, a small uniform \(4 \times 4\) grid of representative registries is sampled for each interface supercell in the fractional in-plane registry space, and the median near-contact gap over these registries is selected. For each trial separation, contacted interfacial atoms are counted separately for the film and substrate rather than counting contact pairs. The selected gap therefore promotes distributed interfacial contact rather than a few highly localized contacts, ensuring that the FGW distance is evaluated for a reasonable initial contact geometry.

After the gap determination, BO is carried out over the in-plane registry space (Fig.~\ref{fig: workflow}c). For a given interface supercell, the two reference bulk structures and each translated interface structure along the BO trajectory are converted to graphs by pymatgen and ASE. In each graph, atoms are represented as nodes, and their local coordination environments are encoded as node features using radial basis functions (RBFs). Pairwise interatomic distances are used as edge attributes, with all atom pairs connected to form a fully connected graph. Specifically, the interface structure is partitioned into film-side and substrate-side regions, from which separate graphs are constructed and compared independently with their corresponding parent bulk graphs using the FGW distance computed with the Python Optimal Transport (POT) library\cite{Flamary2021}. The combined FGW distance, obtained by summing the two-sided FGW contributions, is then used to guide the BO process. A Gaussian process (GP) surrogate model implemented with scikit-learn\cite{Pedregosa2011} is iteratively updated to propose translations towards an optimized registry.

The workflow establishes a reproducible and physically informed protocol for generating initial structures in high-throughput interface screening, supporting the construction of high-quality datasets for machine-learning applications. However, as the central metric of the workflow, the FGW distance requires systematic evaluation in terms of its structural interpretability, relationship with relaxed energies and effectiveness in registry optimization.

\subsection{Structural-energetic landscape comparison}

\begin{figure}[H]
    \centering
    \includegraphics[width=\textwidth]{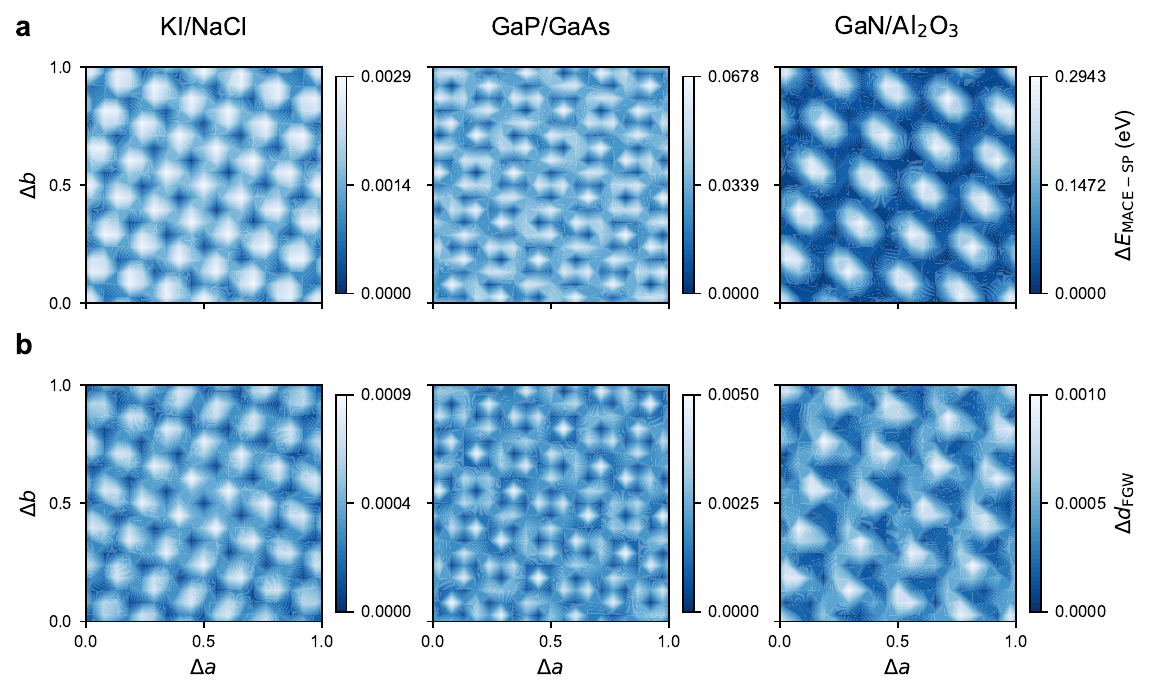}
    \caption{
    Comparison of (a) \(\Delta E_{\mathrm{MACE-SP}}\) and (b) \(\Delta d_{\mathrm{FGW}}\) landscapes for the KI/NaCl, GaP/GaAs and GaN/\(\mathrm{Al_{2}O_{3}}\) interfaces. For each interface, values are referenced to the corresponding landscape minimum.
    }
    \label{fig: landscapes}
\end{figure}

Fig.~\ref{fig: landscapes} compares the relative MACE single-point energy (\(\Delta E_{\mathrm{MACE-SP}}\)) and relative FGW distance (\(\Delta d_{\mathrm{FGW}}\)) landscapes for the KI(001)/NaCl(001), GaP(001)/GaAs(001) and GaN(0001)/\(\mathrm{Al_{2}O_{3}}\)(0001) interfaces. The three representative interfaces are selected to span distinct bonding regimes, assessing the transferability of the FGW distance across chemically diverse interfaces. Stoichiometric KI and NaCl terminations are brought into contact to avoid additional electrostatic effects associated with surface polarity. P-terminated GaP is placed against Ga-terminated GaAs, while Ga-terminated GaN is placed against O-terminated \(\mathrm{Al_{2}O_{3}}\), providing established III-V\cite{Vurgaftman2001} and nitride/oxide\cite{Zhang2019} reference interfaces. Detailed bulk structural data, lattice-matching parameters and initial interfacial geometries are provided in Section S1 of the Supporting Information.

Uniform centred registry grids are used to evaluate the MACE single-point energies and FGW distances shown in Fig.~\ref{fig: landscapes}. A \(19 \times 19\) grid is used for all three interface systems, providing consistent sampling of the two-dimensional registry space. In Fig.~\ref{fig: landscapes}, all landscapes are plotted as relative values referenced to the minimum within each interface system. The colour variation therefore highlights registry-dependent changes rather than differences in absolute offsets.

For unrelaxed interface structures, registry-dependent energies are sensitive to the local chemical and geometric compatibility established across the interface. The FGW distance defined here is designed to quantify this compatibility by using the ideal bulk environments as structural references and comparing the interfacial regions with their corresponding parent phases in a graph-based representation. In the registry searches, the surface orientation, termination, stoichiometry and interfacial separation are fixed, leaving the lateral translation as the primary structural degree of freedom. The MACE single-point energy landscapes in Fig.~\ref{fig: landscapes}a and the FGW distance landscapes in Fig.~\ref{fig: landscapes}b exhibit the same registry-dependent periodicity across all three interface systems. For the GaP/GaAs and GaN/\(\mathrm{Al_{2}O_{3}}\) interfaces, the landscapes also show similar spatial patterns, with broadly aligned low- and high-value regions. For the KI/NaCl interface, however, deviations occur between structural and energetic extrema, reflecting additional chemistry-specific contributions that are not explicitly encoded in the metric. These results indicate that the FGW distance captures local structural compatibility, which constitutes an important component of registry-dependent energy preferences. It provides a physically motivated metric for interface-registry screening.

As shown in Fig.~S4, the MACE single-point energy varies by only around 3 meV across the registry space of the KI/NaCl interface, nearly an order of magnitude smaller than the corrugation observed for the other two interfaces. The corresponding unrelaxed potential-energy surface is essentially flat on the scale of the registry comparison, and the energy corrugation is well below the typical uncertainty of the MACE model\cite{Bilbrey2025}. Nevertheless, systematic model errors may largely cancel since the comparison concerns relative energies among closely related registries within the same interface. When plotted with an individual colour scale (Fig.~\ref{fig: landscapes}a), both quantities indicate the same underlying registry periodicity of the rock-salt/rock-salt interface, but the remaining corrugation may reflect electrostatic effects instead of the structural compatibility measured by FGW distance. More generally, the energy of an unrelaxed structure is evaluated only at the imposed geometry and does not necessarily identify the desired basin of attraction\cite{Wales1992}.

\subsection{Predictive capability for relaxed energy}

\begin{figure}[H]
    \centering
    \includegraphics[width=\textwidth]{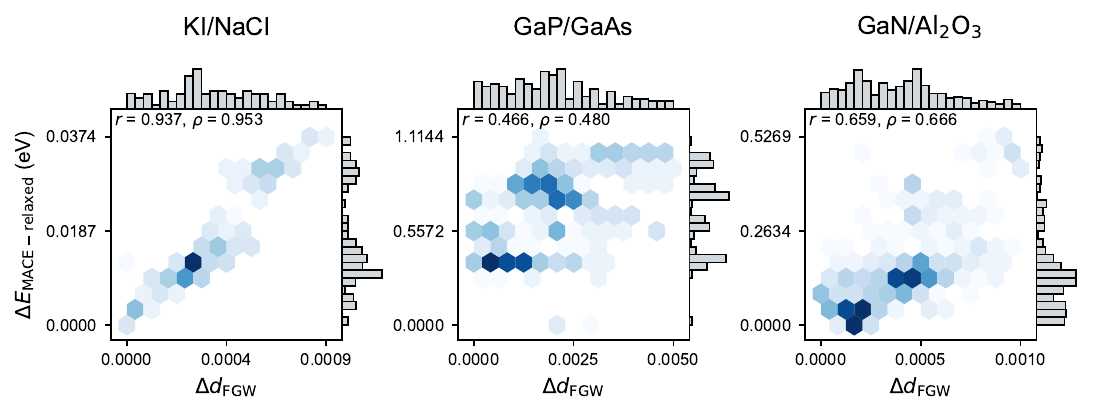}
    \caption{
    Correlation between \(\Delta d_{\mathrm{FGW}}\) and \(\Delta E_{\mathrm{MACE-relaxed}}\) for the KI/NaCl, GaP/GaAs and GaN/\(\mathrm{Al_{2}O_{3}}\) interfaces. Pearson (\(r\)) and Spearman (\(\rho\)) correlation coefficients are indicated in each panel.
    }
    \label{fig: mace_correlation}
\end{figure}

\begin{figure}[H]
    \centering
    \includegraphics[width=\textwidth]{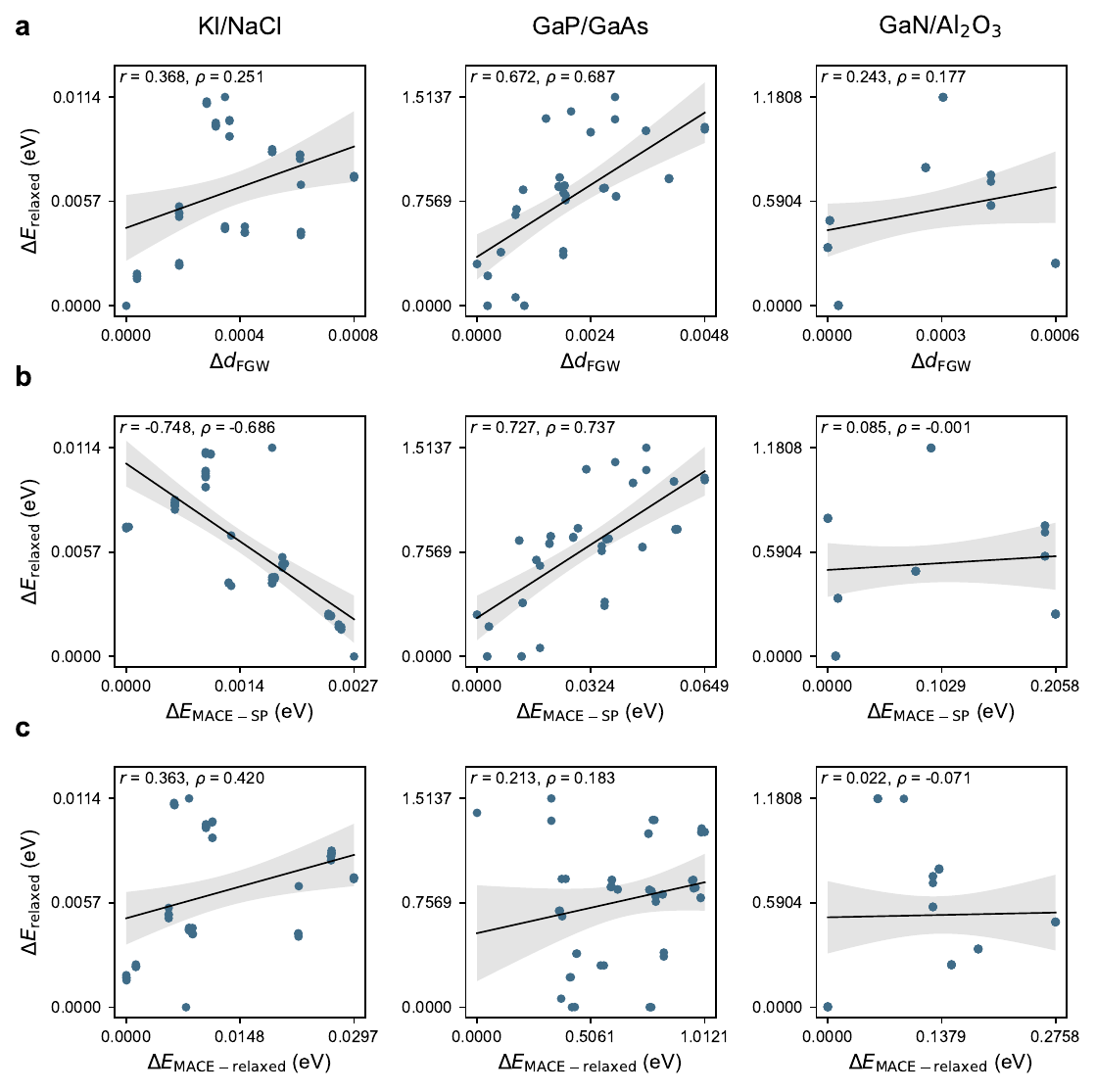}
    \caption{
    Correlations of (a) \(\Delta d_{\mathrm{FGW}}\), (b) \(\Delta E_{\mathrm{MACE-SP}}\), and (c) \(\Delta E_{\mathrm{MACE-relaxed}}\) with \(\Delta E_{\mathrm{relaxed}}\) for the KI/NaCl, GaP/GaAs and GaN/\(\mathrm{Al_{2}O_{3}}\) interfaces. Solid lines show linear regression fits, with shaded regions indicating 95\% confidence intervals. Each panel reports the Pearson (\(r\)) and Spearman (\(\rho\)) correlation coefficients.
    }
    \label{fig: dft_correlation}
\end{figure}

The FGW distance is compared with the relaxed energy to evaluate the heuristic that bulk-like continuity identifies favourable starting structures. Fig.~\ref{fig: mace_correlation} compares relative FGW distance with relative MACE-relaxed energy (\(\Delta E_{\mathrm{MACE-relaxed}}\)) and Fig.~\ref{fig: dft_correlation}a compares relative FGW distance with relative DFT-relaxed energy (\(\Delta E_{\mathrm{relaxed}}\)) for the KI/NaCl, GaP/GaAs and GaN/\(\mathrm{Al_{2}O_{3}}\) interfaces. The FGW distances and energies are referenced to their respective system-specific minima to facilitate comparison.

MACE-relaxed energies are evaluated on the denser \(19 \times 19\) grid described above. Owing to the computational cost and convergence challenges associated with DFT interface relaxation, DFT-relaxed energies are evaluated on a uniform centred \(7 \times 7\) grid for each interface. The grid provides systematic coverage while keeping the number of DFT calculations tractable. As in the landscape calculations, the use of an odd centred grid is retained to reduce symmetry-induced degeneracy among the sampled registries.

The correlations in Fig.~\ref{fig: mace_correlation} and \ref{fig: dft_correlation}a show that low-FGW registries tend to relax to lower-energy structures. This indicates that the metric preferentially identifies structurally favourable initial configurations and provides an effective basis for pre-relaxation screening. For the KI/NaCl interface, the MACE-relaxed energy exhibits an exceptionally strong correlation with the FGW distance, with Pearson (\(r\)) and Spearman rank (\(\rho\)) correlation coefficients of 0.937 and 0.953, respectively (Fig.~\ref{fig: mace_correlation}). In contrast, the corresponding DFT correlation decreases markedly to \(r = 0.368\) and \(\rho = 0.251\) in Fig.~\ref{fig: dft_correlation}a. Although the variation in MACE-relaxed energy, approximately 40 meV, is significantly larger than the \(\sim3\) meV single-point corrugation in Fig.~\ref{fig: landscapes}a, it remains small in absolute terms. At this energy scale, the strong MACE-FGW correlation may partially reflect the comparative smooth variation in the energy landscape predicted by the pretrained MLIP\cite{Deng2025}. However, the DFT-relaxed energy may additionally capture subtle registry-dependent electrostatic effects that are not directly described by the structure-based FGW metric, potentially weakening the FGW-DFT correlation.

The GaP/GaAs interface shows broadly comparable positive correlations following MACE and DFT relaxation, with \(r = 0.466\) and \(\rho = 0.480\) for the MACE-relaxed energy and \(r = 0.672\) and \(\rho=0.687\) for the DFT-relaxed energy. By contrast, the GaN/\(\mathrm{Al_{2}O_{3}}\) interface shows a decrease from \(r = 0.659\) and \(\rho = 0.666\) for MACE to \(r = 0.243\) and \(\rho = 0.177\) for DFT. In this case, the \(7 \times 7\) grid contains significant symmetry-induced redundancy and provides only a few structurally distinct configurations. As a result, the FGW-DFT correlation is not statistically robust. Furthermore, to evaluate the sensitivity of these trends to the FGW representation, Section S3 of the Supporting Information examines the effects of \(\alpha\) and the element embedding on the correlations. Across all three interfaces, varying \(\alpha\) produces only minor changes in the correlations between FGW distance and either MACE-relaxed or DFT-relaxed energy. The choice of element embedding likewise has a limited influence for GaP/GaAs and GaN/\(\mathrm{Al_{2}O_{3}}\). Nevertheless, for KI/NaCl, MEGNet yields the strongest correlations among the embeddings, reflecting its broader encoding of energetically relevant elemental information. In addition, the marginal histograms in Fig.~\ref{fig: mace_correlation} illustrate that the correlations extend across the sampled distributions rather than being dominated by a few extreme configurations. The greater concentration of GaN/\(\mathrm{Al_{2}O_{3}}\) registries near the energy minimum is consistent with a broader low-energy basin.

While the absolute correlation values may appear modest in some of these cases, we note that the lowest FGW distance interface is consistently in the top 4 most stable symmetry-inequivalent interfaces from DFT. This underscores the utility of the metric for screening studies, a point further explored in Section~\ref{FGW-BO}.

The correlations of MACE single-point and MACE-relaxed energies with DFT-relaxed energies are shown in Fig~\ref{fig: dft_correlation}b and \ref{fig: dft_correlation}c. Overall, the FGW heuristic provides a more consistent and reliable relationship to the lowest energy DFT structures than either MACE screening approaches. This is particularly evident for the GaN/\(\mathrm{Al_{2}O_{3}}\) interface, where neither MACE single-point energy nor MACE-relaxed energy shows an appreciable correlation with DFT-relaxed energy. How these correlation trends translate into the prioritization of low-energy registries is further examined under limited screening budgets.

\subsection{Performance of FGW-guided screening\label{FGW-BO}}

\begin{figure}[H]
    \centering
    \includegraphics[width=\textwidth]{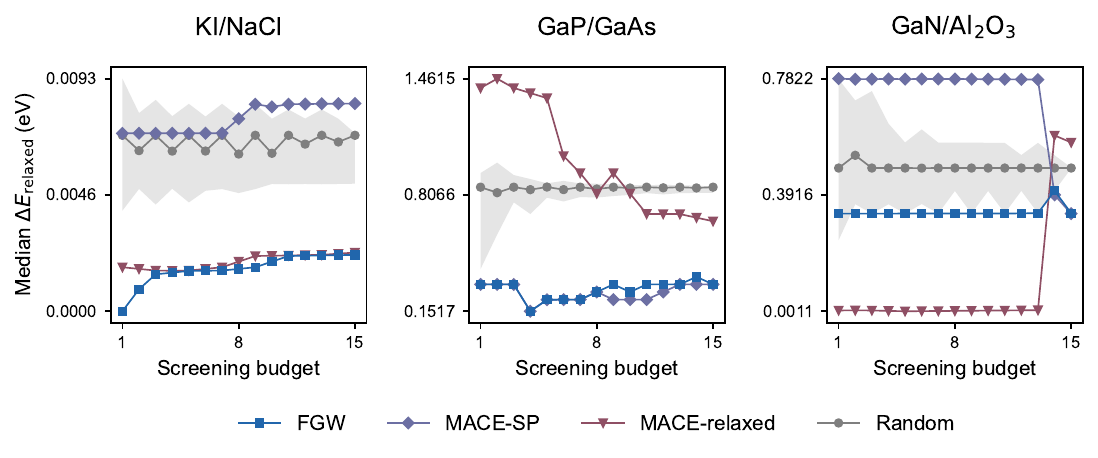}
    \caption{
    Screening performance of FGW distance, MACE single-point energy (MACE-SP), MACE-relaxed energy and random selection for the KI/NaCl, GaP/GaAs and GaN/\(\mathrm{Al_{2}O_{3}}\) interfaces. For each deterministic screening method, the median \(\Delta E_{\mathrm{relaxed}}\) of the selected registries is reported at each screening budget. For random selection, the solid and shaded region denote the median and interquartile range, respectively, of the median \(\Delta E_{\mathrm{relaxed}}\) obtained over 1000 independent trials.
    }
    \label{fig: performance}
\end{figure}

In order to evaluate the practical utility of the FGW distance in a realistic screening scenario, FGW-guided selection is benchmarked against selection based on MACE single-point and MACE-relaxed energies and random selection (Fig.~\ref{fig: performance}). The purpose is to test whether the initial structures selected by the energy-independent FGW distance provide more favourable starting points for subsequent high-accuracy relaxation. 

The benchmark is performed on the same uniform centred \(7 \times 7\) grid described above. At each screening budget, the registries are ranked according to the corresponding screening metric, and performance is evaluated using the median \(\Delta E_{\mathrm{relaxed}}\) of the selected registries. For each interface system, \(\Delta E_{\mathrm{relaxed}}\) is defined relative to the minimum relaxed energy on the grid. Random selection is repeated over 1000 independent trials, with the median and interquartile range reported. The use of median values reduces sensitivity to individual outliers, since relaxation outcomes can vary substantially across registry space. Such variation is consistent with the local and path-dependent nature of geometry optimization on complex potential-energy surfaces\cite{Wales1998, Wales2018}.

As shown in Fig.~\ref{fig: performance}, FGW-guided selection outperforms random selection across all three interfaces. For KI/NaCl, selection based on MACE-relaxed energy performs comparably to FGW-guided selection, whereas MACE single-point-energy selection performs worse than random selection. For GaP/GaAs, the opposite trend is observed, with MACE single-point-energy selection performing similarly to FGW-guided selection and MACE-relaxed-energy selection underperforming random selection. For GaN/\(\mathrm{Al_{2}O_{3}}\), MACE-relaxed-energy selection outperforms FGW-guided selection at screening budgets below 14, but its performance deteriorates rapidly as the budget increases. The symmetry-reduced results in Fig.~S5 indicate that this early advantage arises from the repeated selection of symmetry-equivalent low-energy registries. MACE single-point-energy selection again performs worse than random selection. Collectively, the FGW distance should not be interpreted as a guarantee of locating the global minimum, but as an effective prescreening metric that biases the search towards stable, energetically favourable regions.

 In addition, the BO convergence results shown in Fig.~S6, demonstrate that the adopted BO strategy generally identifies low-FGW regions more efficiently than random sampling. BO therefore serves as an effective search component within the proposed FGW-guided workflow.

\section{Conclusions}

We have introduced a new heuristic metric for automatic generation of physically and chemically motivated, atomistic models of heterointerfaces. Our method uses a combination of graph networks and optimal transport theory, based on a fused Gromov Wasserstein (FGW) distance, to encode the intuition that the most stable interfaces are those that best replicate the bulk structure of the constituent materials and can be seen as a generalisation of coincidence site lattices to handle heterointerfaces. We test and demonstrate our method on three diverse systems covering ionic interfaces, covalent interfaces and mixed ionic/covalent interfaces. We compare our method to low energy structure identified by exhaustive searches with both density functional theory (DFT) and modern machine learned interatomic potentials (MLIPs). We find reasonable correlation (both rank correlation and value correlation) with the values from MLIPs, with the lowest energy DFT structure consistently amongst the most favourable identified by our metric across all systems and levels of theory compared. We further demonstrate that our metric can be used with Bayesian optimisation (BO) to efficiently search complex atomistic structure space. Across all systems considered our metric offers consistent improvement against random search, in marked contrast to MLIP based BO, where different search setups dramatically affect  the result. This study therefore demonstrates the utility of FGW as a lightweight, reliable encoding of chemical intuition, suitable for integration into high-throughput searching of heterointerface structures. The code required for applying FGW is made available as an easily installable Python package.

\section{Methods}

\subsection{Zur-McGill lattice matching}

All symmetrically distinct Miller indices up to 1 are generated for the film and substrate bulk materials. Lattice matching is then performed for each resulting film-substrate orientation pair using the Zur-McGill algorithm as implemented in pymatgen. For a pair of interface-parallel two-dimensional primitive surface lattices with area \(A_1\) and \(A_2\), possible common superlattices are identified by searching for positive integer superlattice orders \(r_1\) and \(r_2\) that satisfy

\[
r_{1}A_{1} \simeq r_{2}A_{2}
\]

where \(r_1\) and \(r_2\) define the superlattice areas \(r_{1}A_{1}\) and \(r_{2}A_{2}\), respectively. Candidate superlattice cells are reduced to canonical lattice representations and identified as matches when the relative mismatches in superlattice area, in-plane lattice-vector lengths, and intervector angle are within the prescribed tolerances. The area, length and angle tolerances are set to 6\%, 3\% and 2\%, respectively. Furthermore, to control the size of the matched interface supercells for computational tractability, the maximum allowed matching area is set to \(150~\text{\AA}^2\) for KI/NaCl and GaP/GaAs, and to \(80~\text{\AA}^2\) for GaN/\(\mathrm{Al_{2}O_{3}}\).

For each lattice-matched slab pair, the film and substrate slabs are subsequently flipped independently to enumerate all combinations of interface-facing terminations. The lattice-matching step considers only geometric commensurability of the two surface lattices and does not assess their chemical compatibility. Before registry screening, the target interface supercells are selected from the constructed candidates.

\subsection{Near-contact gap selection}

To ensure that interface registries are compared under physically meaningful contact conditions, a near-contact separation is determined for each candidate interface supercell. This step prevents unrealistically large or small interfacial separations from dominating the FGW evaluation. The interfacial registry is parametrized by fractional in-plane translation of the film slab relative to the fixed substrate slab along the two in-plane lattice vectors, \(\mathbf{x} = (\Delta a, \Delta b) \in [0, 1)^2\). A pair of atoms \(i\) and \(j\) on opposite sides of the interface is defined as being in contact when

\[
d_{ij}(\mathbf{x})
\le
r_i^{\mathrm{cov}} + r_j^{\mathrm{cov}} + \tau
\]

where \(d_{ij}(\mathbf{x})\) is the cross-interface distance, \(r_i^{\mathrm{cov}}\) and \(r_j^{\mathrm{cov}}\) are the element-specific covalent radii, and \(\tau\) is a small distance tolerance. Cross-interface distances are evaluated using periodic boundary conditions in the two in-plane directions, as implemented using pymatgen and ASE.

In order to avoid selecting a gap that is biased by a single registry, the near-contact condition is evaluated over a uniform \(4 \times 4\) registry grid, and the median near-contact gap is used for each interface system. For each trial separation, contacts are counted separately on the film and substrate sides. An interface is considered to reach the near-contact condition when the smaller of the two side-resolved counts is at least 3. For larger interface supercells, where 3 atoms may not be sufficient to represent spatially distributed interfacial contact, the criterion is instead set to 5\% of the atom count on the side of the interface containing fewer atoms. Based on this protocol, the near-contact gaps are set to \(3.6~\text{\AA}\) for KI/NaCl, \(2.3~\text{\AA}\) for GaP/GaAs and \(1.9~\text{\AA}\) for GaN/\(\mathrm{Al_{2}O_{3}}\).

\subsection{FGW distance calculation}

To calculate FGW distances, interfacial registries and bulk references are converted into graphs using pymatgen and ASE. Each graph is represented by atom-wise node features describing the local chemical environments, together with an intra-graph structural distance matrix encoding pairwise geometric relations between atoms.

For a graph containing \(N\) atoms, the node feature \(\mathbf{h}_{i}\) of atom \(i\) is constructed from a radial-basis expansion of its neighbouring atoms.

\[
\mathbf{h}_{i} 
=
\sum_{j \in \mathcal{N}_{i}}
f_{\mathrm{c}}(d_{ij})
[
\phi_1(d_{ij}), \ldots, \phi_K(d_{ij}) 
]
\otimes \mathbf{e}_{Z_j}
\]

where \(\mathcal{N}_{i}\) is the set of neighbouring atoms within the cutoff radius \(r_{\mathrm{cut}}\), \(d_{ij}\) is the interatomic distance between atoms \(i\) and \(j\), \(f_{\mathrm{c}}\) is a cosine cutoff function, and \(\mathbf{e}_{Z_j}\) is the element embedding vector of neighbouring atom \(j\). The \(k\)-th radial basis function \(\phi_{k}\) is defined as

\[
\phi_k(d_{ij})
=
\exp[-\gamma(d_{ij} - \mu_k)^2]
\]

where \(\mu_k\) is the centre of the \(k\)-th radial basis function and \(K\) is the number of radial basis functions. The RBF centres are distributed uniformly between 0 and \(r_{\mathrm{cut}}\), using 3 radial basis centres per \(\text{\AA}\). The Gaussian width is determined by the spacing between adjacent RBF centres, \(\Delta r = r_{\mathrm{cut}}/(K-1)\), with \(\gamma = 1/\Delta r^2\). \(r_{\mathrm{cut}}\) is set to the larger of the maximum third-neighbour distances of the two parent bulk phases, ensuring adequate coverage of the short-range coordination environments of both sides of the interface.

MEGNet element embeddings\cite{Chen2019} are used for all interface systems to incorporate periodic chemical trends learned from crystal formation-energy prediction. The FGW distance is interpreted as a within-system relative metric, rather than as an absolute measure across different interface systems.

The structural distance matrix \(\mathbf{C}\) is constructed from all pairwise interatomic distances under a two-dimensional minimum-image convention.

\[
C_{ij}
=
\min_{n_1, n_2 \in \mathbb{Z}}
\left\|
\mathbf{r}_{i}-\mathbf{r}_{j}
+
n_{1}\mathbf{a}_{1}
+
n_{2}\mathbf{a}_{2}
\right\| 
\]

where \(\mathbf{r}_{i}\) and \(\mathbf{r}_{j}\) are the Cartesian coordinates of atoms \(i\) and \(j\), \(\mathbf{a}_{1}\) and \(\mathbf{a}_{2}\) are the two in-plane lattice vectors, and \(n_{1}\) and \(n_{2}\) are integer lattice translations. This definition gives the shortest periodic interatomic distance along the two in-plane periodic directions, while minimum-image wrapping is not applied along the surface-normal direction.

For each registry, the interface structure is partitioned into film-side and substrate-side regions, from which two separate interface-region graphs are constructed and compared with the corresponding bulk reference graphs. Because each bulk graph uses the same atomic subset and geometry as its interface-region counterpart, the paired graphs have identical structural distance matrices by construction. The side-resolved scheme avoids directly comparing the full heterogeneous interface with a single bulk phase, which would otherwise introduce a large registry-independent contribution from the intrinsic difference between the two bulk phases. Although the graph geometry is fixed across registries, the node features encode cross-interface neighbours and thus respond to lateral registry shifts.

The dissimilarity between two graphs is measured using the FGW distance, implemented with the POT library.

\[
d_{\mathrm{FGW}, \alpha}(G_1, G_2)
=
\min_{\mathbf{T} \in \Pi(\mathbf{p}, \mathbf{q})}
(1 - \alpha)\langle\mathbf{T}, \mathbf{M}\rangle
+
\alpha\sum_{i,k,j,l}L(C_{1, ik}, C_{2, jl})T_{ij}T_{kl}
\]

where \(\mathbf{T}\) is the transport plan between the nodes of the two graphs, \(\mathbf{p}\) and \(\mathbf{q}\) are the node-weight distributions, \(\mathbf{M}\) is the node-feature cost matrix, and \(\mathbf{C}_{1}\) and \(\mathbf{C}_{2}\) are the intra-graph structural distance matrices. \(\mathbf{p}\) and \(\mathbf{q}\) are taken as uniform over the graph nodes. The first term compares node features, whereas the second term compares pairwise relational structures within the two graphs. The balance between these two contributions is controlled by \(\alpha\), and \(L\) is the squared-loss function. The feature cost matrix and structural distance matrices are normalized to comparable numerical scales, and \(\alpha\) is set to 0.5 to incorporate both feature and structural contributions. Since the structural term inherited from Gromov-Wasserstein (GW) formulation leads to a non-convex quadratic optimization problem, the numerical solution is sensitive to the initial transport plan\cite{Takeda2025}. To reduce sensitivity to local minima, each FGW calculation is repeated with 80 randomized initial transport plans, and the minimum value obtained across these runs is retained. The combined FGW distance is defined as the sum of the film-side and substrate-side FGW distances, quantifying the total registry-induced deviation from bulk environments on both sides of the interface.

\subsection{Bayesian optimization}

Bayesian optimization is performed over the two-dimensional in-plane registry space of each interface. The interfacial registry is parametrized by fractional film translation \(\mathbf{x} = (\Delta a, \Delta b) \in [0, 1) \times [0, 1)\). The out-of-plane separation is fixed at the predetermined near-contact gap. For each trial registry, geometric validity is first checked based on cross-interface interatomic distances. Registries with excessively short or insufficient interfacial contacts are excluded from Gaussian-process fitting. For valid registries, the combined FGW distance is evaluated and used as the objective to be minimized.

The optimization is initialized with 8 registries generated using a scrambled Sobol low-discrepancy sequence. A Gaussian process surrogate model is implemented using scikit-learn and fitted to the valid sampled points. The model uses a constant-scaled Mat\'ern kernel with \(\nu = 2.5\). Periodicity is incorporated by mapping the fractional registry coordinates to 

\[
\tilde{\mathbf{x}}
= 
\left[
\cos(2\pi\Delta a),
\sin(2\pi\Delta a),
\cos(2\pi\Delta b),
\sin(2\pi\Delta b)
\right]
\]

before fitting. Target values are normalized and a diagonal regularization term of \(10^{-6}\) is added for numerical stability. At each refinement step, 4096 candidate registries are generated uniformly across the registry space. The next evaluation point is selected by maximizing the expected improvement (EI) acquisition function for minimization of the FGW distance.

\[
\mathrm{EI}(\mathbf{x}) 
= 
(f_{\mathrm{min}} - \mu(\mathbf{x}) - \xi)\Phi(z) 
+
\sigma(\mathbf{x})\phi(z)
\]

\[
z
=
\frac{f_{\mathrm{min}} - \mu(\mathbf{x}) - \xi}{\sigma(\mathbf{x})}
\]

where \(\mu(\mathbf{x})\) and \(\sigma(\mathbf{x})\) are the Gaussian-process predictive mean and standard deviation, \(f_{\mathrm{min}}\) is the minimum FGW distance observed so far, \(\xi = 10^{-4}\) is the exploration parameter, \(\Phi\) and \(\phi\) denote the cumulative distribution function and the probability density function, respectively, of the standard normal distribution. The optimization is repeated for 50 refinement steps. For a screening budgets of \(k\), the \(k\) registries with the lowest FGW distances are selected for further evaluation.

\subsection{MLIP calculations}

MLIP calculations are performed using the pre-trained MACE-MP medium model in single precision through the ASE interface, without additional system-specific tuning. The calculations provide an intermediate high-throughput validation layer for assessing registry-dependent energy trends at a scale inaccessible to exhaustive DFT calculations. However, many foundation MLIPs rely on finite-cutoff atomic representations and do not explicitly model all long-range electrostatic, polarization and charge-transfer contributions\cite{McKenna2016, Cheng2025, Kim2025}. DFT calculations are therefore used as the final high-fidelity validation.

Single-point calculations are carried out for the unrelaxed interfacial structures sampled on the registry grid. For MLIP relaxations, the supercell lattice vectors are kept fixed, and all atomic positions are allowed to relax. The unconstrained relaxation protocol avoids constraint-induced discontinuities at boundaries between fixed and mobile atoms, which may be underrepresented in the near-equilibrium pretraining data of universal MLIPs\cite{Deng2025}. Geometry optimizations are performed using the FIRE optimizer until the maximum force falls below \(0.03~\text{eV/\AA}\), or until a maximum of 300 optimization steps is reached.

\subsection{DFT calculations}

DFT calculations are performed using the Vienna Ab initio Simulation Package (VASP)\cite{Kresse1996}. The projector augmented-wave (PAW) method is employed with the Perdew-Burke-Ernzerhof (PBE) exchange-correlation functional within the generalized gradient approximation (GGA)\cite{Kresse1999}. A plane-wave energy cutoff of 520 eV is used. The Brillouin zone is sampled using \(\Gamma\)-centred k-point meshes with a spacing of \(0.25~\text{\AA}^{-1}\). Dipole corrections are applied along the surface-normal direction to reduce artificial electrostatic interactions between periodic images.

For geometry optimizations, the supercell lattice vectors are kept fixed, and only atomic positions are relaxed. Atoms in the half of each slab farthest from the interface are constrained to preserve bulk-like reference regions, while the remaining near-interface atoms are allowed to relax. The electronic energy convergence criterion is set to \(10^{-5}\) eV, and ionic relaxation is continued until the residual forces are below \(0.03~\text{eV/\AA}\). Static single-point calculations are then performed on the relaxed structures using a tighter energy convergence criterion of \(10^{-6}\) eV.

\section{Data availability}

The processed data supporting this study are available on GitHub at \url{https://github.com/YuxuanTang2002/RegFGW}. The raw DFT data are available on Zenodo at \url{https://doi.org/10.5281/zenodo.21764050}.

\section{Code availability}

The source code and analysis scripts used in this study are available at \url{https://github.com/YuxuanTang2002/RegFGW}.

\section{Acknowledgements}

We are grateful to the UK Materials and Molecular Modelling Hub for computational resources, which is partially funded by EPSRC (EP/T022213/1, EP/W032260/1 and EP/P020194/1). KTB acknowledges funding from UKRI (EP/Y014405/1, EP/Y000552/1, EP/Y028775/1, EP/Y028759/1).

\section{Author contributions}

Y.T. conducted calculations, analysed the data, and prepared the original draft. K.T.B. conceived the idea, edited and reviewed the manuscript, and supervised the project.

\section{Competing interests}

The authors declare no competing interests.

\bibliographystyle{unsrt}
\bibliography{reference}

\end{document}


\maketitle

\section{Bulk structural data and interface lattice matching}

\begin{table}[htbp]
    \centering
    \caption{
    Bulk structures and conventional lattice parameters used for interface construction, obtained from the Materials Project database\cite{Jain2013}.
    }
    \label{table: bulk}
    \begin{tabular}{lllll}
    \toprule
    Material &
    MP ID &
    Crystal structure &
    Space group &
    \begin{tabular}[c]{@{}l@{}}
    Conventional lattice \\
    parameters
    \end{tabular} \\
    \midrule
    KI &
    mp-22898 &
    rock salt &
    \(\mathrm{Fm}\bar{3}\mathrm{m}\) &
    \(a = 7.08~\text{\AA}\) \\
    NaCl &
    mp-22862 &
    rock salt &
    \(\mathrm{Fm}\bar{3}\mathrm{m}\) &
    \(a = 5.59~\text{\AA}\) \\
    GaP &
    mp-2490 &
    zinc blende &
    \(\mathrm{F}\bar{4}\mathrm{3m}\) &
    \(a = 5.45~\text{\AA}\) \\
    GaAs &
    mp-2534 &
    zinc blende &
    \(\mathrm{F}\bar{4}\mathrm{3m}\) &
    \(a = 5.75~\text{\AA}\) \\
    GaN &
    mp-804 &
    wurtzite &
    \(\mathrm{P6}_{3}\mathrm{mc}\) &
    \(a = 3.19~\text{\AA},\;c = 5.19~\text{\AA}\) \\
    \(\mathrm{Al_{2}O_{3}}\) &
    mp-1143 &
    corundum &
    \(\mathrm{R}\bar{3}\mathrm{c}\) &
    \(a = 4.81~\text{\AA},\;c = 13.12~\text{\AA}\) \\
    \bottomrule
    \end{tabular}
\end{table}

\begin{table}[htbp]
    \centering
    \caption{
    Construction details of the coherent interface models. Interface labels, orientations and terminations are reported in the film/substrate order. All models satisfy the area, length and angle tolerances of 6\%, 3\% and 2\%, respectively.
    }
    \label{table: inteface}
    \begin{tabular}{llll}
    \toprule
    Interface &
    Orientation pair &
    Termination pair &
    Interface area (\(\text{\AA}^2\)) \\
    \midrule
    KI/NaCl &
    (001)/(001) &
    stoichiometric/stoichiometric &
    124.91 \\
    GaP/GaAs &
    (001)/(001) &
    P-terminated/Ga-terminated &
    148.79 \\
    GaN/\(\mathrm{Al_{2}O_{3}}\) &
    (0001)/(0001) &
    Ga-terminated/O-terminated &
    59.99 \\
    \bottomrule
    \end{tabular}
\end{table}

\begin{figure}[htbp]
    \centering
    \includegraphics[page=1, width=\textwidth, trim=0cm 6cm 0cm 6cm, clip]{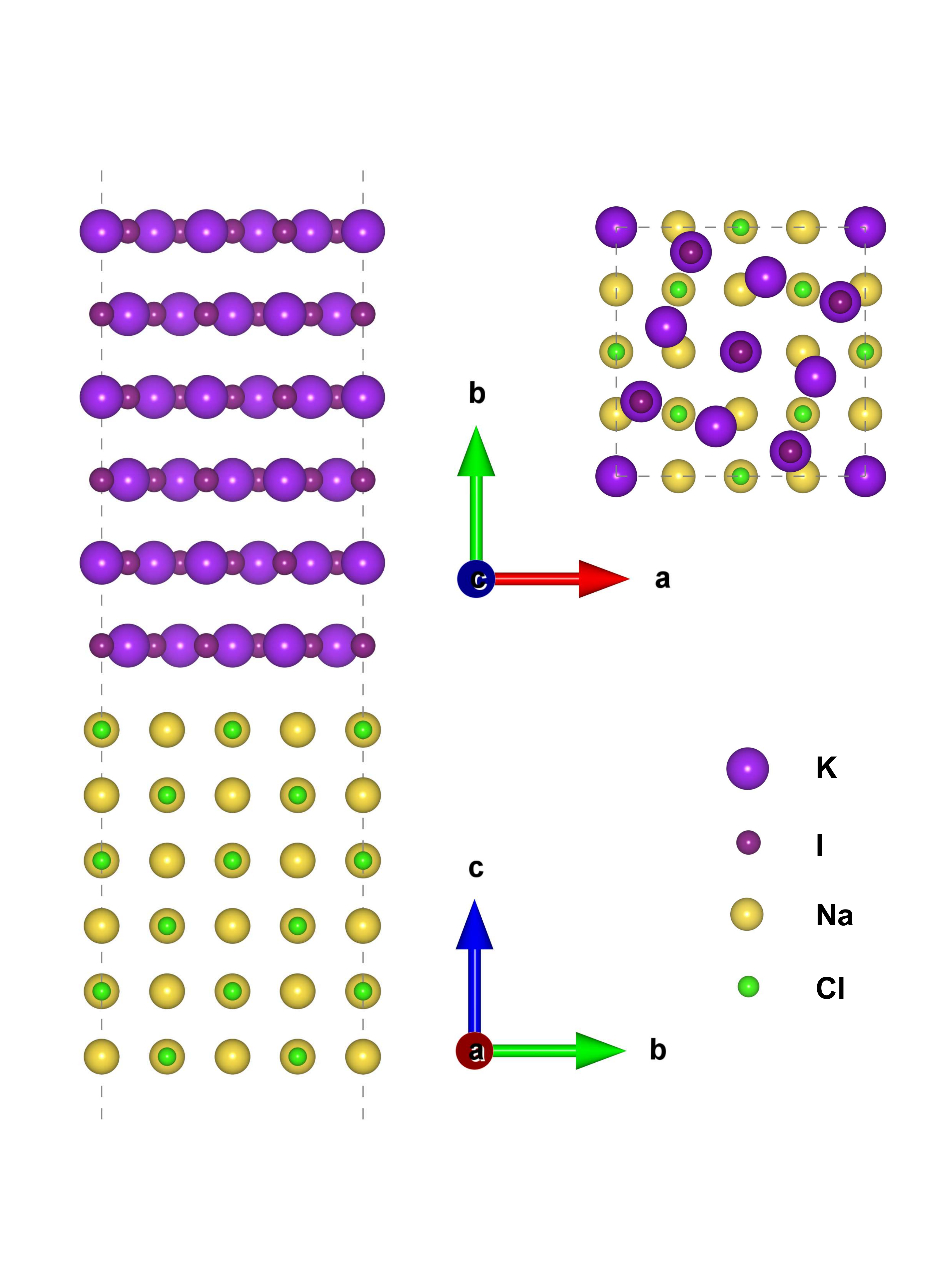}
    \caption{
    Atomic structure of the KI/NaCl interface at the near-contact interfacial separation. Both side and top views are shown.
    }
    \label{fig: KI/NaCl}
\end{figure}

\begin{figure}[htbp]
    \centering
    \includegraphics[page=2, width=\textwidth, trim=0cm 9cm 0cm 9cm, clip]{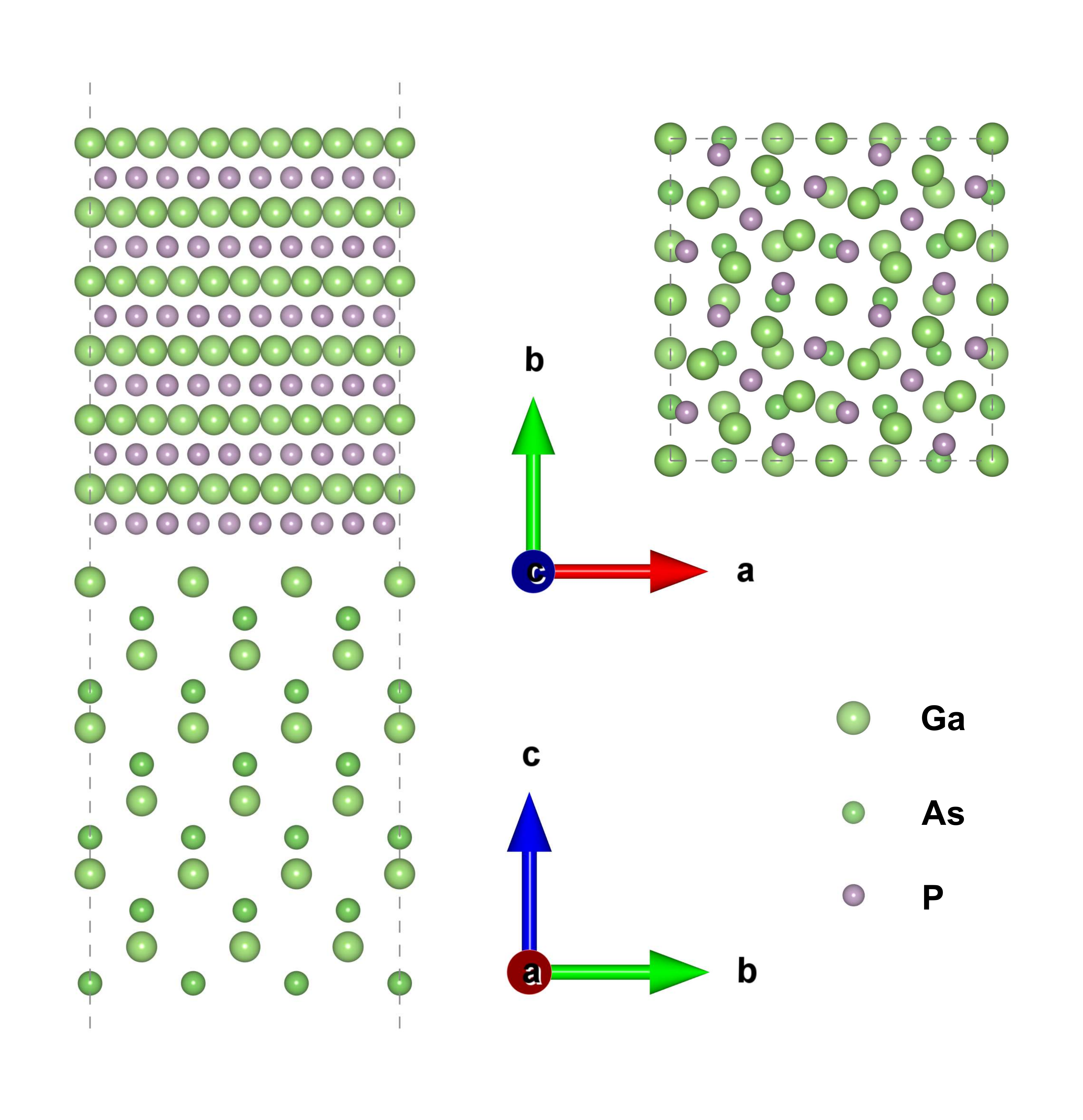}
    \caption{
    Atomic structure of the GaP/GaAs interface at the near-contact interfacial separation. Both side and top views are shown.
    }
    \label{fig: GaP/GaAs}
\end{figure}

\begin{figure}[htbp]
    \centering
    \includegraphics[page=3, width=\textwidth, trim=0cm 12cm 0cm 12cm, clip]{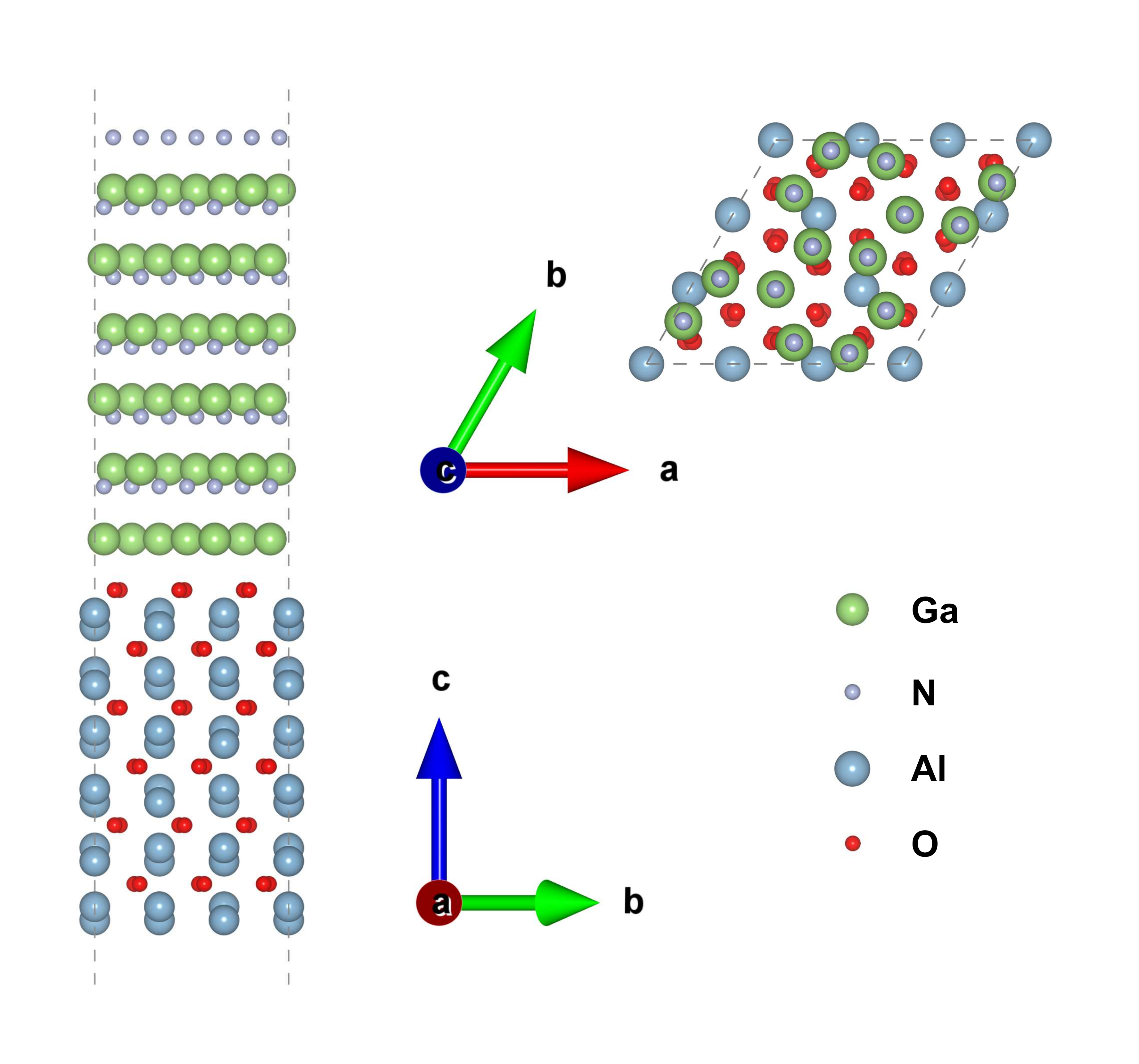}
    \caption{
    Atomic structure of the GaN/\(\mathrm{Al_{2}O_{3}}\) interface at the near-contact interfacial separation. Both side and top views are shown.
    }
    \label{fig: GaN/Al2O3}
\end{figure}

\section{Structural-energetic landscapes with common colour scales}

\begin{figure}[H]
    \centering
    \includegraphics[width=\textwidth]{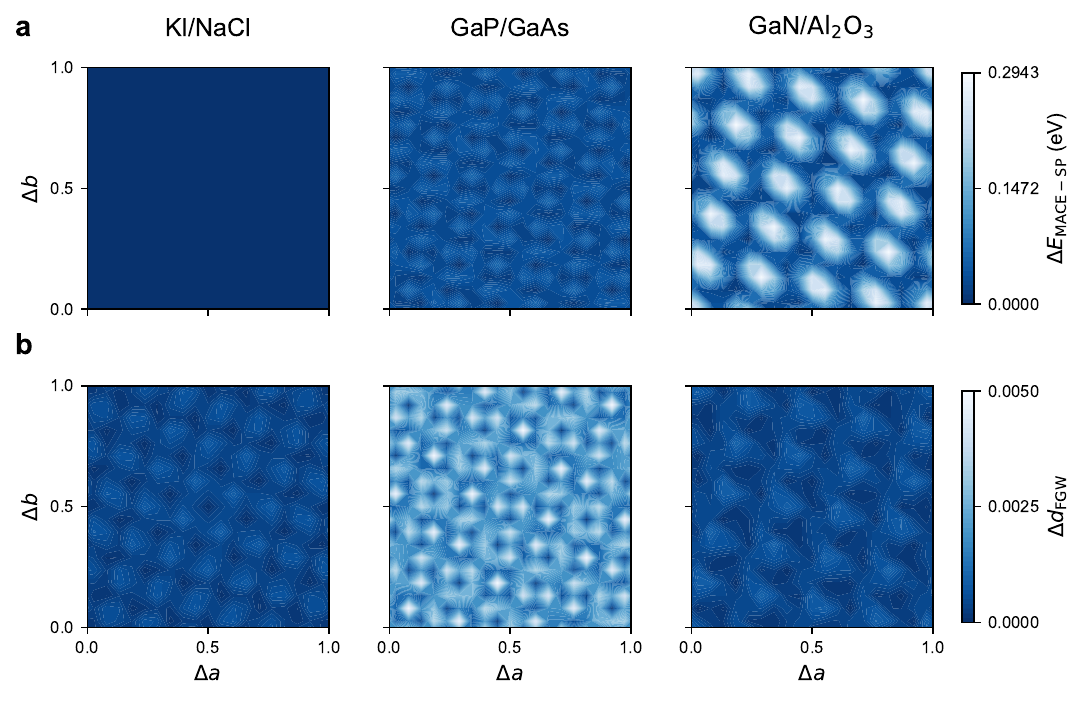}
    \caption{
    Comparison of (a) relative MACE single-point energy (\(\Delta E_{\mathrm{MACE-SP}}\)) and (b) relative FGW distance (\(\Delta d_{\mathrm{FGW}}\)) landscapes for the KI/NaCl, GaP/GaAs and GaN/\(\mathrm{Al_{2}O_{3}}\) interfaces. For each interface, values are referenced to the corresponding landscape minimum. A common colour scale is used for all panels within each row.
    }
    \label{fig: common_scale_landscapes}
\end{figure}

\section{FGW sensitivity analysis}

\begin{table}[htbp]
    \centering
    \caption{
    Sensitivity of the correlations to the FGW mixing parameter, \(\alpha\), for the KI/NaCl, GaP/GaAs, GaN/\(\mathrm{Al_{2}O_{3}}\) interfaces. Values of 0.25, 0.50 and 0.75 are compared using the MEGNet\cite{Chen2019} element embedding. Pearson (\(r\)) and Spearman (\(\rho\)) correlation coefficients are reported for \(\Delta d_{\mathrm{FGW}}\) with relative MACE-relaxed energies (\(\Delta E_{\mathrm{MACE-relaxed}}\)) and relative DFT-relaxed energies (\(\Delta E_{\mathrm{relaxed}}\)).
    }
    \label{table: alpha_sensitivity}
    \begin{tabular}{llllll}
    \toprule
    System & 
    \(\alpha\) & 
    \(r(\Delta E_{\mathrm{MACE-relaxed}})\) &
    \(\rho(\Delta E_{\mathrm{MACE-relaxed}})\) &
    \(r(\Delta E_{\mathrm{relaxed}})\) &
    \(\rho(\Delta E_{\mathrm{relaxed}})\) \\
    \midrule
    KI/NaCl & 0.25 & 
    0.937 & 0.953 & 0.368 & 0.251 \\
    KI/NaCl & 0.50 &
    0.937 & 0.953 & 0.368 & 0.251 \\
    KI/NaCl & 0.75 &
    0.937 & 0.953 & 0.368 & 0.251 \\
    GaP/GaAs & 0.25 &
    0.466 & 0.480 & 0.672 & 0.687 \\
    GaP/GaAs & 0.50 &
    0.466 & 0.480 & 0.672 & 0.687 \\
    GaP/GaAs & 0.75 &
    0.466 & 0.480 & 0.672 & 0.687 \\
    GaN/\(\mathrm{Al_{2}O_{3}}\) & 0.25 &
    0.590 & 0.664 & 0.243 & 0.177 \\
    GaN/\(\mathrm{Al_{2}O_{3}}\) & 0.50 &
    0.659 & 0.666 & 0.243 & 0.177 \\
    GaN/\(\mathrm{Al_{2}O_{3}}\) & 0.75 &
    0.659 & 0.666 & -0.002 & 0.168 \\    
    \bottomrule
    \end{tabular}
\end{table}

\begin{table}[htbp]
    \centering
    \caption{
    Sensitivity of the correlations to the element embeddings for the KI/NaCl, GaP/GaAs and GaN/\(\mathrm{Al_{2}O_{3}}\) interfaces. MEGNet, Magpie\cite{Ward2016} and CGNF\cite{Jang2024} embeddings are compared with \(\alpha\) set to 0.5. \(r\) and \(\rho\) are reported for \(\Delta d_{\mathrm{FGW}}\) with \(\Delta E_{\mathrm{MACE-relaxed}}\) and \(\Delta E_{\mathrm{relaxed}}\).
    }
    \label{table: embedding_sensitivity}
    \begin{tabular}{llllll}
    \toprule
    System &
    Embedding &
    \(r(\Delta E_{\mathrm{MACE-relaxed}})\) &
    \(\rho(\Delta E_{\mathrm{MACE-relaxed}})\) &
    \(r(\Delta E_{\mathrm{relaxed}})\) &
    \(\rho(\Delta E_{\mathrm{relaxed}})\) \\
    \midrule
    KI/NaCl & MEGNet &
    0.937 & 0.953 & 0.368 & 0.251 \\
    KI/NaCl & CGNF &
    0.324 & 0.173 & -0.616 & -0.698 \\
    KI/NaCl & Magpie &
    0.486 & 0.290 & -0.441 & -0.593 \\
    GaP/GaAs & MEGNet &
    0.466 & 0.480 & 0.672 & 0.687 \\
    GaP/GaAs & CGNF &
    0.474 & 0.491 & 0.686 & 0.682 \\
    GaP/GaAs & Magpie &
    0.469 & 0.488 & 0.683 & 0.673 \\
    GaN/\(\mathrm{Al_{2}O_{3}}\) & MEGNet &
    0.659 & 0.666 & 0.243 & 0.177 \\
    GaN/\(\mathrm{Al_{2}O_{3}}\) & CGNF &
    0.657 & 0.700 & 0.242 & 0.354 \\
    GaN/\(\mathrm{Al_{2}O_{3}}\) & Magpie &
    0.643 & 0.698 & 0.276 & 0.354 \\
    \bottomrule
    \end{tabular}
\end{table}

\section{Screening performance after structural deduplication}

\begin{figure}[H]
    \centering
    \includegraphics[width=\textwidth]{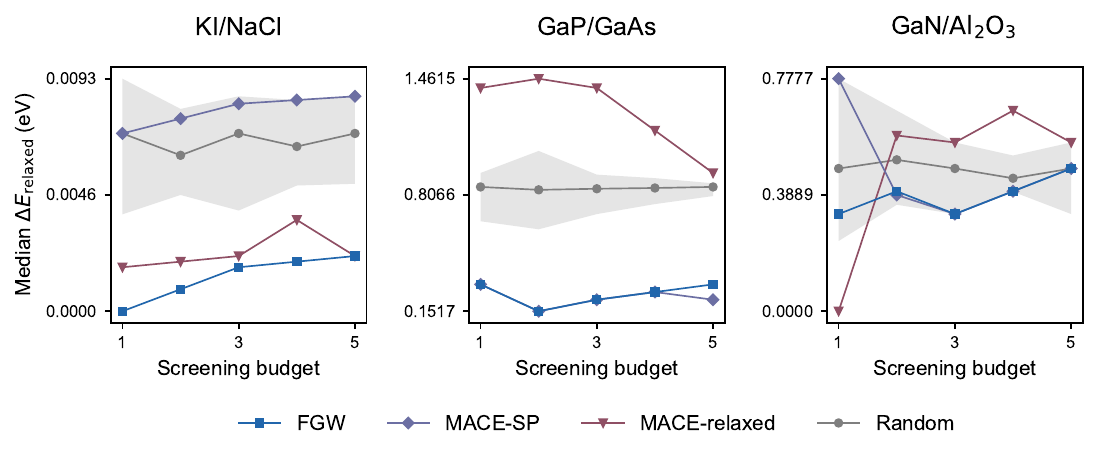}
    \caption{Screening performance of FGW distance, MACE single-point energy (MACE-SP), MACE-relaxed energy (MACE-relaxed) and random search after structure deduplication for the KI/NaCl, GaP/GaAs and GaN/\(\mathrm{Al_{2}O_{3}}\) interfaces. Performance is evaluated using the median \(\Delta E_{\mathrm{relaxed}}\) of the candidates selected for each screening budget. For random search, the line and shaded region denote the median and interquartile range over 1000 repeated selections, respectively.}
    \label{fig: performance_unique}
\end{figure}

\section{Bayesian optimization convergence}

\begin{figure}[H]
    \centering
    \includegraphics[width=\textwidth]{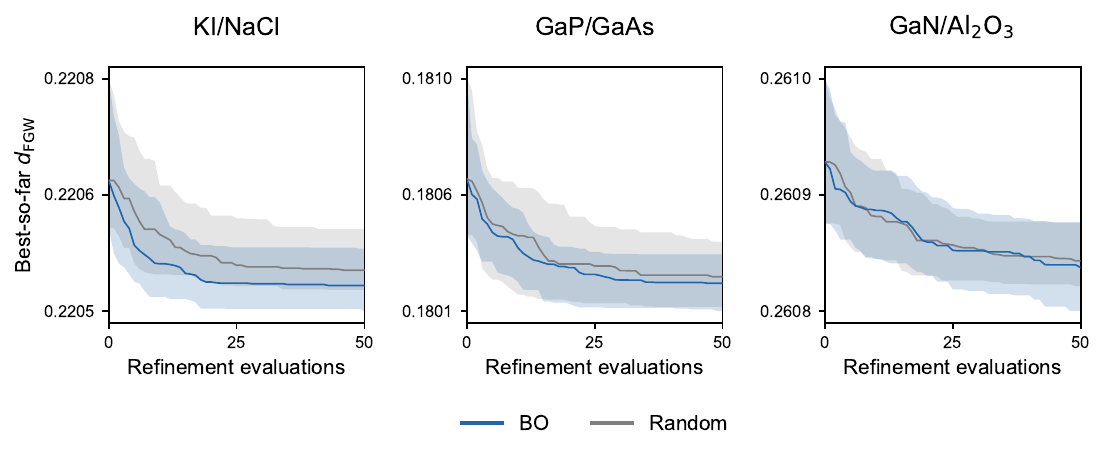}
    \caption{Convergence of Bayesian optimization (BO) and random search for minimizing \(d_{\mathrm{FGW}}\) across the registry spaces of the KI/NaCl, GaP/GaAs and GaN/\(\mathrm{Al_{2}O_{3}}\) interfaces. Curves show the median best-so-far \(d_\mathrm{FGW}\) over 80 independent trials, and shaded regions show the interquartile ranges.}
    \label{fig: bo_convergence}
\end{figure}

\bibliographystyle{unsrt}
\bibliography{reference}